\documentclass[twocolumn,showpacs,preprintnumbers,amsmath,amssymb]{revtex4}
\usepackage{graphicx}% Include figure files
\usepackage{dcolumn}% Align table columns on decimal point
\usepackage{bm}% bold math

\usepackage{subfig}
\usepackage{adjustbox}
\usepackage{lpic}

\usepackage{braket}
\usepackage{tensor}

\newcommand{\hc}{\textrm{hc}}

\newcommand{\ee}{\text{e}}
\newcommand{\ii}{\text{i}}

\newcommand{\hn}{\hat{n}}
\newcommand{\hhm}{\hat{m}}
\newcommand{\hB}{\hat{B}}
\newcommand{\hC}{\hat{C}}
\newcommand{\hE}{\hat{E}}
\newcommand{\hI}{\hat{I}}
\newcommand{\hxi}{\hat{\xi}}
\newcommand{\hzeta}{\hat{\zeta}}

\newcommand{\EinD}{\hE^{{\text{in}}\dagger}}
\newcommand{\Ein}{\hE^\text{in}}
\newcommand{\Eout}{\hE^\text{out}}

\newcommand{\Iin}{\hI^\text{in}}
\newcommand{\Iout}{\hI^\text{out}}

\newcommand{\dd}[1]{\!{\text{d} #1} \,}

\usepackage[greek,english]{babel} % english = default
\usepackage[LGR,T1]{fontenc}
\usepackage{lmodern,xspace} %and amsmath
\def\PI{\ensuremath{\text{\foreignlanguage{greek}{p}}}\xspace}

\begin{document}

%\preprint{APS/123-QED}

\title{Phase-noise limitations on single-photon cross-phase modulation\\ with differing group velocities}

\author{Justin Dove}
\email{dove@mit.edu}
\author{Christopher Chudzicki}
\author{Jeffrey H. Shapiro}%
\affiliation{%
Research Laboratory of Electronics, Massachusetts Institute of
Technology, Cambridge, Massachusetts 02139, USA
}%

\date{\today}% It is always \today, today,

\begin{abstract}
  A framework is established for evaluating {\sc cphase} gates that
  use single-photon cross-phase modulation (XPM) originating from the
  Kerr nonlinearity. Prior work [Phys.~Rev.~A {\bf 73,} 062305
  (2006)], which assumed that the control and target pulses propagated
  at the same group velocity, showed that the causality-induced phase
  noise required by a non-instantaneous XPM response function
  precluded the possibility of high-fidelity $\PI$-radian conditional
  phase shifts. The framework presented herein incorporates the more
  realistic case of group-velocity disparity between the control and
  target pulses, as employed in existing XPM-based fiber-optical
  switches. Nevertheless, the causality-induced phase noise identified
  in [Phys.~Rev.~A {\bf 73,} 062305 (2006)] still rules out
  high-fidelity $\PI$-radian conditional phase shifts. This is shown
  to be so for both a reasonable theoretical model for the XPM
  response function and for the experimentally-measured XPM response
  function of silica-core fiber.
\end{abstract}

\maketitle

\section{Introduction}
Optics-based quantum computing is an attractive possibility.
Single-photon source and detector technologies are rapidly maturing
\cite{0034-4885-75-12-126503, doi:10.1021/nl404400d, Calkins:13,
  1408.1124}, enabling robust photonic-qubit creation and detection.
Moreover, single-qubit gates are easily realized with linear optics,
and photons are the inevitable carriers for the long-distance
entanglement distribution needed to network quantum computers.
However, optics-based quantum computing is not without its Achilles'
heel, namely the extremely challenging task of realizing a
high-fidelity, deterministic, two-qubit entangling gate, such as the
{\sc cphase} gate.

Knill, Laflamme, and Milburn~\cite{Knill2001} proposed a solution to
the preceding two-qubit gate problem by exploiting the nonlinearity
afforded by photodetection in conjunction with the introduction of
ancilla photons. Their scheme is intrinsically probabilistic, so it
requires high-efficiency adaptive measurement techniques and large
quantities of ancilla photons to realize useful levels of quantum
computation. Consequently, it remains prudent to continue research on
more traditional approaches to all-optical two-qubit gates. A prime
example is the nonlinear-optical approach first suggested by Chuang
and Yamamoto~\cite{PhysRevA.52.3489}, who proposed using Kerr-effect
cross-phase modulation (XPM) to impart a $\PI$-radian phase shift on a
single-photon pulse, conditioned on the presence of another
single-photon pulse.

The fact that the Chuang--Yamamoto architecture provides a
deterministic all-optical universal gate set for quantum computation
continues to spur work on highly-nonlinear optical fibers
\cite{edamatsu, gaeta}, but the single-photon level has yet to be
reached. Chuang and Yamamoto's analysis treated the control and target
as single-spatiotemporal-mode fields. Later work
\cite{PhysRevA.73.062305}, however, examined their architecture using
continuous-time XPM theory. Dismissing the possibility of an
instantaneous XPM response---owing to its failure to reproduce
experimentally-observed classical results---it showed that a causal,
non-instantaneous response function introduces fidelity-degrading
phase noise which precludes constructing a high-fidelity {\sc cphase}
gate. That analysis assumed control and target pulses propagating at
the same group velocity, which implied that a uniform single-photon
phase shift could not be realized in the fast-response regime, wherein
those pulses have durations much longer than that of the XPM response
function. Yet fast-response XPM \emph{is} used for imparting uniform
conditional phase shifts in fiber-optical switching with classical
control pulses \cite{974167}. Those switches' pulses have different
group velocities, so that one propagates through the other within the
XPM medium.

In this paper we develop a continuous-time quantum XPM theory for
pulses with differing group velocities \cite{footnote3}, and then use
it to assess the feasibility of extending the fiber-switching
technique to the single-photon regime for creating a {\sc cphase}
gate. We show that causality-induced phase noise still rules out
high-fidelity $\PI$-radian conditional phase shifts for both a
reasonable theoretical model for the XPM response function and for the
experimentally-measured XPM response function of silica-core fiber.

\section{Quantum XPM Theory}
Our theory begins with classical XPM for a pair of single-spatial-mode
continuous-time scalar fields---with center frequencies $\omega_A$ and
$\omega_B$ and complex envelopes $E_A(z,t)$ and $E_B(z,t)$---that
propagate from $z=0$ to $z=L$ through an XPM medium. Because we are
interested in ultimate limits on the utility of XPM for two-qubit
gates, we neglect loss, dispersion, and self-phase modulation. Thus
the behavior of the classical complex envelopes of interest is
governed by the coupled-mode equations \cite{agrawal2001nonlinear}:
\begin{subequations}
\label{eq:coupledmode}
\begin{align}
\left( \frac{\partial}{\partial{z}} + \frac{1}{v_A} \frac{\partial}{\partial{t}} \right) E_{A}(z,t) &= \ii n_A(z,t) E_A(z,t), \\
\left( \frac{\partial}{\partial{z}} + \frac{1}{v_B} \frac{\partial}{\partial{t}} \right) E_{B}(z,t) &= \ii n_B(z,t) E_B(z,t).
\end{align}
\end{subequations}
Here, $v_A$ and $v_B$, satisfying $v_B > v_A$, are the group
velocities of $E_A(z,t)$ and $E_B(z,t)$, and $n_A(z,t)$ and $n_B(z,t)$
are the intensity-dependent refractive indices that these fields
encounter. For convenient linking to the quantum analysis, we
normalize $E_A(z,t)$ and $E_B(z,t)$ to make $\hbar\omega_{K}I_{K}(z,t)
=\hbar\omega_{K}|E_{K}(z,t)|^2,$ for $K= A,B$, the powers carried by
these fields. The nonlinear refractive indices are then given by
\cite{footnote4}
\begin{subequations}
\begin{align}
n_A(z,t) &= \eta \int_{-\infty}^t \dd{t'} h(t-t') I_B(z,t'),  \\
n_B(z,t) &= \eta \int_{-\infty}^t \dd{t'} h(t-t') I_A(z,t'),
\end{align}
\end{subequations}
where $\eta$ is the strength of the nonlinearity and $h(t)$ is its
real-valued, causal response function, normalized to satisfy
$\int_0^\infty \dd{t} h(t) = 1$.

In the quantum theory for the preceding XPM setup, $E_A(z,t)$ and
$E_B(z,t)$ become baseband field operators, $\hE_A(z,t)$ and
$\hE_B(z,t)$ with units $\sqrt{\rm photons/s}$. At the input and
output planes, $z=0$ and $z=L$, these field operators must satisfy the
canonical commutation relations for free fields, viz.,
\begin{subequations}
\label{commutators}
\begin{align}
  [\hE_K(z,t),\hE_J(z,s)] &= 0,\\
  [\hE_K(z,t),\hE_J^\dagger(z,s)] &= \delta_{JK}\delta(t-s),
\end{align}
\end{subequations}
for $K = A,B$, $J = A,B$, and $z=0,L$.
Unless $h(t)=\delta(t)$, which \cite{PhysRevA.73.062305} has ruled out for its failure to reproduce experimentally-observed classical results, Langevin noise terms must be added to the classical coupled-mode equations to ensure that the output fields have the required commutators.  Here, we take a cue from the work
of Boivin \em et al\/\rm.~\cite{PhysRevLett.73.240}, which developed a
continuous-time quantum theory of self-phase modulation and which \cite{PhysRevA.73.062305} extended to XPM when both fields have the same group velocity.  The
quantum coupled-mode equations that result are
\begin{subequations}
\label{eq:quantumcoupledmode}
\begin{align}
\left( \frac{\partial}{\partial{z}} + \frac{1}{v_A}
  \frac{\partial}{\partial{t}} \right) \hE_{A}(z,t) &= i [\hn_A(z,t)
+ \hhm_A(z,t)] \hE_A(z,t), \\
\left( \frac{\partial}{\partial{z}} + \frac{1}{v_B}
  \frac{\partial}{\partial{t}} \right) \hE_{B}(z,t) &= i [\hn_B(z,t)
+ \hhm_A(z,t)] \hE_B(z,t).
\end{align}
\end{subequations}

In terms of the photon-flux operators
$\hI_{K}(z,t)\equiv \hE_{K}^\dagger(z,t)\hE_{K}(z,t)$, for $K= A,B$, the nonlinear refractive indices are now operator-valued and given by
\begin{subequations}
\begin{align}
\hn_A(z,t) &= \eta \int_{-\infty}^t \dd{t'} h(t-t') \hI_B(z,t'),  \\
\hn_B(z,t) &= \eta \int_{-\infty}^t \dd{t'} h(t-t') \hI_A(z,t').
\end{align}
\end{subequations}
The Langevin noise operators $\hhm_A(z,t)$ and $\hhm_B(z,t)$ are 
\begin{subequations}
\begin{eqnarray}
\lefteqn{\hhm_A(z,t) = \int_{0}^\infty \frac{\dd{\omega}}{2\PI}
\sqrt{\eta H_\text{im}(\omega)} } \nonumber \\ && \times\,\{[\hB(z,\omega) - \ii
\hC^\dagger(z,\omega)] \ee^{-\ii\omega t} + \hc\},  \\[.05in]
\lefteqn{\hhm_B(z,t) = \int_{0}^\infty \frac{\dd{\omega}}{2\PI}
\sqrt{\eta H_\text{im}(\omega)} } \nonumber \\ 
&& \times\,\{[\hB(z,\omega) + \ii
\hC^\dagger(z,\omega)] \ee^{-\ii\omega t} + \hc\},
\end{eqnarray}
\end{subequations}
where $H_\text{im}(\omega)$ is the imaginary part of the frequency
response $H(\omega) = \int_0^\infty \dd{t} h(t) e^{i \omega t}$,
$\hB(z,\omega)$ and $\hC(z,\omega)$ are independent frequency-domain
bosonic field operators \cite{footnote1} taken to be in thermal states
at absolute temperature $T$, and $\hc$ denotes the Hermitian
conjugate.

Equation~(\ref{eq:quantumcoupledmode}) can be solved to yield the following input-output relations
\begin{subequations}\label{eq:io}
\begin{align}
  \Eout_A(t) &= \text{e}^{\text{i}\hxi_A(t)}
  \text{e}^{\text{i}\hzeta_A(t)} \Ein_A(t),
  \\
  \Eout_B(t) &= \text{e}^{\text{i}\hxi_B(t)}
  \text{e}^{\text{i}\hzeta_B(t)} \Ein_B(t)\text{,}
\end{align}
\end{subequations}
for the output field operators, $\Eout_K(t)\equiv\hE_K(L,t+L/v_K)$, in terms of the input field operators, 
 $\Ein_K(t)\equiv\hE_K(0,t)$, the \emph{phase-shift operators} \cite{footnote2}
\begin{subequations}
\begin{align}
  \hzeta_A(t) \equiv \eta \int_0^{L} \dd{z} \int \dd{s} h(t-s) \Iin_B(s+z/u), \\
  \hzeta_B(t) \equiv \eta \int_{0}^{L} \dd{z} \int \dd{s}
  h(t-s) \Iin_A(s-z/u),
\end{align}
\end{subequations}
where $\Iin_K(t) \equiv {\EinD_K}(t)\Ein_K(t)$ and $1/u \equiv 1/v_A - 1/v_B$,
and the \emph{phase-noise operators} 
\begin{subequations}
\begin{eqnarray}
  \lefteqn{\hxi_A(t) \equiv \int_0^L\dd{z}\int_{0}^{\infty} \frac{\dd{\omega}}{2\pi} \sqrt{\eta H_\text{im}(\omega) }} \nonumber \\
  && \times\,\{ [\hB(z,\omega) - i \hC^{\dagger}(z,\omega)]e^{-i \omega (t+z/v_A)} + \hc\}, \\[.05in]
  \lefteqn{\hxi_B(t) \equiv \int_0^L\dd{z}\int_{0}^{\infty}  \frac{\dd{\omega}}{2\pi}
  \sqrt{\eta H_\text{im}(\omega) }} \nonumber \\ 
  && \times\,\{ [\hB(z,\omega) + i
  \hC^{\dagger}(z,\omega)]e^{-i \omega (t+z/v_B)} + \hc\}.
\end{eqnarray}
\end{subequations}
These input-output relations ensure that $\Eout_A(t)$ and
$\Eout_B(t)$ have the proper free-field commutators, as required by Eq.~(\ref{commutators}). 

The phase-noise
operators have nonzero commutator
\begin{eqnarray}
  \lefteqn{[\hxi_A(t) , \hxi_B(s) ] = } \nonumber \\
  && \ \ii \eta \int_0^{L} \dd{z} [h(s-t-z/u) - h(t-s+z/u)],
\end{eqnarray}
and they are in a zero-mean jointly Gaussian
state that is characterized by the symmetrized autocorrelation functions
\begin{equation}
\braket{\hxi_K(t)\hxi_K(s) + \hxi_K(s)\hxi_K(t)} =    
\int\frac{\dd{\omega}}{\pi} S_{\xi\xi}(\omega)\cos[\omega(t-s)]\text{,}
\label{eq:xi3}
\end{equation}
for $K=A,B$, with spectrum
\begin{equation}\label{eq:xi4}
S_{\xi\xi}(\omega) = \eta H_\text{im}(\omega)
\coth\left(\frac{\hbar\omega}{2k_BT}\right)\text{,}
\end{equation}
where $k_B$ is the Boltzmann constant. For the theory to make physical
sense, it must be that $H_\text{im}(\omega) \geq 0$ for all $\omega\ge
0$~\cite{Shapiro:97, PhysRevLett.73.240}, because noise spectra must
be nonnegative.

\section{XPM-Based CPHASE Gate}
To build a {\sc cphase} gate from the preceding quantum XPM
interaction we proceed as follows. Consistent with dual-rail logic
\cite{PhysRevA.52.3489}, the input and output field operators are
chosen to be in states in the Hilbert space spanned by their
computational basis states, $\{|0\rangle_K, |1\rangle_K : K = A,B\}$.
We will take $\ket{0}_K$ to be the vacuum state, and set
\begin{align}
\ket{1}_K = \int \dd{t} \psi_K(t) \ket{t}_K\text{,}
\end{align}
where the wave functions $\{\psi_K(t) : K = A,B\}$ are normalized
($\int \dd{t} |\psi_K(t)|^2 = 1$), and $\ket{t}_K$ is the state of
$\Ein_K(t)$ or $\Eout_K(t)$ in which there is a single photon at time
$t$ and none at all other times. To enforce the interchangeability of
the control and target qubits, we take the single-photon pulses in
each field to have the same pulse shape. Moreover, because we have
assumed $v_B > v_A$, we will assume that $\psi_B(t) = \psi_A(t -
t_d)$, where $t_d > 0$ is a delay, specified below, chosen to allow
the single-photon excitation in $\Ein_B(t)$ to propagate through the
one in $\Ein_A(t)$ while both are within the nonlinear medium, thus
ensuring each imposes a uniform phase shift on the other.

Sufficient conditions for guaranteeing a uniform phase shift are
intuitive and easily derived. Ignoring the phase noise for now, the
phase shifts induced on each field by the presence of a single-photon
pulse in the other field are found by taking the partial trace of the
phase-shift operator for one field with respect to the other:
\begin{subequations}
\begin{align}
\tensor[_B]{\braket{1 | \ee^{\ii\hzeta_A(t)} | 1}}{_B} = & \int
\dd{s} \ee^{\ii \eta\int^L_0\dd{z} h(t-s+z/u)}
|\psi_B(s)|^2, \label{eq:induced-phase-shift_a}\\
\tensor[_A]{\braket{1 | \ee^{\ii\hzeta_B(t)} | 1}}{_A} = & \int
\dd{s} \ee^{\ii \eta\int^L_0\dd{z} h(t-s-z/u)} |\psi_A(s)|^2.
\label{eq:induced-phase-shift_b}
\end{align}
\end{subequations}
From these expressions it is clear that a sufficient condition for a
uniform phase shift on $\Eout_K(t)$ is that the response-function
integrals, i.e., the integrals in the exponents in
Eqs.~(\ref{eq:induced-phase-shift_a}) and (\ref{eq:induced-phase-shift_b}), encapsulate the entirety of the
response function for all times $t$ and $s$ for which $\psi_K(t)$ and
$\psi_J(s)$ are nonzero, where $J \neq K$. Suppose that $h(t)$,
$\psi_A(t)$, and $\psi_B(t)$ are only nonzero over the intervals
$[0,t_h]$, $[-t_\psi/2, t_\psi/2]$, and $[-t_\psi/2 + t_d, t_\psi/2 +
t_d]$, respectively. Although these conditions might not be satisfied
exactly, e.g., when the response function and/or the pulse shapes do
not have bounded support, we can at least take $t_h$ and $t_\psi$ to
represent the nominal durations over which each function is
significantly different from zero. In terms of these time durations,
our sufficient conditions for uniform phase shifts are
\begin{subequations}
\begin{align}
t_d & \geq t_\psi + t_h \label{eq:uniform-phase-1_a}\\
\frac{L}{u} & \geq t_\psi + t_h + t_d. \label{eq:uniform-phase-1_b}
\end{align}
\end{subequations}
Under these conditions, we have that
\begin{align}
\int^L_0\dd{z} h(t-s+z/u) = \int^L_0\dd{z} h(t-s-z/u) = u,
\end{align}
in Eqs.~(\ref{eq:induced-phase-shift_a}) and (\ref{eq:induced-phase-shift_b}), respectively,
due to the normalization of $h(t)$.  Hence 
these equations reduce to
\begin{align}
\tensor[_B]{\braket{1 | \ee^{\ii\hzeta_A(t)} | 1}}{_B} =
\tensor[_A]{\braket{1 | \ee^{\ii\hzeta_B(t)} | 1}}{_A} = \ee^{\ii \eta u},
\end{align}
after making use of the wave functions' normalization.
From this result we see that the uniform phase shift imposed by the presence
of a single-photon excitation is
\begin{align}\label{eq:phase-magnitude}
\phi = \eta u,
\end{align}
so that the strength, $\eta$, of the XPM nonlinearity must be $\eta =
\PI/u$ to realize the desired {\sc cphase} gate

Physically, the condition in Eq.~(\ref{eq:uniform-phase-1_a}) implies
that the fast pulse, $\hE_B(z,t)$, does not enter
the XPM medium until the entirety of the slow pulse, $\hE_A(z,t)$, and
its nonlinear response have propagated into it.  Similarly, Eq.~(\ref{eq:uniform-phase-1_b}) implies that the
slow pulse, $\hE_A(z,t)$, does not exit the XPM
medium until the entirety of the fast pulse, $\hE_B(z,t)$, and its
nonlinear response have propagated out of it. Taking both the delay
and the XPM medium to be as short as possible, which will prove most
favorable with regards to the phase noise, these conditions simplify
to
\begin{subequations}\label{eq:uniform-phase-2}
\begin{align}
t_d & = t_\psi + t_h \\
\frac{L}{u} & = 2(t_\psi + t_h).
\end{align}
\end{subequations}

\section{Vacuum and Single-Photon Fidelities}
A complete fidelity analysis for the preceding XPM-based {\sc cphase} gate would evaluate the overlap between the actual two-field output state from
the XPM interaction and the two-field output state from an ideal {\sc
  cphase} gate, averaged uniformly over all possible two-field input
states.  We, however, will limit our attention to the
vacuum and single-photon fidelities, introduced in \cite{PhysRevA.73.062305}. Let $\Ein_B(t)$, regarded as the gate's control field, be in its vacuum (single-photon) state. The ensuing vacuum (single-photon) fidelity $F_0$ ($F_1$) is the
overlap between the actual state for $\Eout_A(t)$ and that field's ideal state, averaged uniformly over all $\Ein_A(t)$ states on the Bloch sphere. The phase-noise limits we will find for these two fidelities are closely related, so let us begin with $\Ein_B(t)$ being in its vacuum state, in which case the
ideal {\sc cphase}-gate state for $\Eout_A(t)$ is its input state.  The formula from  \cite{PhysRevA.73.062305} for the vacuum fidelity is
\begin{eqnarray} 
  \lefteqn{F_0 =  \frac{1}{3} \left [1 + {\rm Re}\left( \int \dd{t} \braket{\ee^{\ii\hxi_A(t)}}
    |\psi_A(t)|^2 \right) \right.} \nonumber \\[.05in]
    &&\hspace*{-.1in}+\,\left. \int \dd{t} \int \dd{s} |\psi_A(t)|^2 |\psi_A(s)|^2
  \braket{\ee^{\ii[\hxi_A(t) -\hxi_A(s)]}} \right].
  \label{eq:vac-fid}
\end{eqnarray}
We can place an upper bound on the vacuum fidelity by letting
$T=0\,\text{K}$, to minimize the phase noise, and setting the third
term in Eq.~(\ref{eq:vac-fid}) to 1, to maximize its value. Using the
$\braket{e^{i\hxi_A(t)}}$ value for the phase noise's zero-mean
Gaussian state whose spectrum is given by Eq.~(\ref{eq:xi4}) with
$T=0$\,K, we find that
\begin{align}
F_0 \le \frac{2}{3} + \frac{1}{3}
\exp\!\left(-\frac{\eta L}{4\PI} \int \dd{\omega}|H_\text{im}(\omega)|\right).
\label{nonuniformF0bound}
\end{align}
Under our uniform-phase-shift conditions,
Eqs.~(\ref{eq:uniform-phase-2}) and (\ref{eq:phase-magnitude}), this
bound on the vacuum fidelity becomes
\begin{align}
\label{fidelitybound}
F_0 \le \frac{2}{3} + \frac{1}{3}
\exp\!\left(-\frac{\phi}{2\PI}(t_\psi + t_h) \int \dd{\omega}|H_\text{im}(\omega)|\right).
\end{align}

It is readily apparent from (\ref{fidelitybound}) that the vacuum
fidelity decreases as the phase shift increases. Likewise, it is clear
that perfect fidelity for a nonzero phase shift is impossible,
\emph{even in theory}, for any physically-valid response function.
Perfect fidelity for a nonzero phase-shift $\phi$ requires either
$t_\psi+t_h = 0$ or $|H_\text{im}(\omega)| = 0$ for all $\omega$. The
former is impossible for non-instantaneous pulse shapes and response
functions, while the latter is impossible for non-instantaneous,
causal response functions. An even looser, more favorable bound can be
gotten by presuming operation to be in the slow-response regime,
wherein $t_\psi \ll t_h$. For a $\PI$-radian phase shift, we are then
left with
\begin{align}
F_0 \le F_\text{max} \equiv \frac{2}{3} + \frac{1}{3}
\exp\!\left(-(t_h/2) \int \dd{\omega}|H_\text{im}(\omega)|\right).
\label{eq:vac-fid-max}
\end{align}

This $F_\text{max}$ result also applies to the single-photon fidelity,
whose general expression is \cite{PhysRevA.73.062305},
\begin{align} 
 F_1 &=  \frac{1}{3} \left [1 + {\rm Re}\left( \ee^{-i\phi}\int \dd{t} \braket{\ee^{\ii\hxi_A(t)}}\braket{\ee^{\ii\hzeta_A(t)}}
    |\psi_A(t)|^2 \right) \right. \nonumber \\[.05in]
    &+\,\int \dd{t} \int \dd{s} |\psi_A(t)|^2 |\psi_A(s)|^2 \braket{\ee^{\ii[\hxi_A(t) -\hxi_A(s)]}} \nonumber \\[.05in]
  &\times \left.\braket{\ee^{\ii[\hzeta_A(t)-\hzeta_A(s)]}} \right],
  \label{eq:singlephoton-fid}
\end{align}
which reduces to the result in Eq.~(\ref{eq:vac-fid}) when the XPM
interaction produces a uniform $\phi$-radian phase shift. Assuming
$\phi = \PI$ and operation in the slow-response regime, we then get
\begin{align}
F_1 \le \frac{2}{3} + \frac{1}{3}
\exp\!\left(-(t_h/2) \int \dd{\omega}|H_\text{im}(\omega)|\right)
\label{eq:singlephoton-fid-max}
\end{align}
from Eq.~(\ref{eq:singlephoton-fid}), thus putting the same optimistic
but likely unobtainable upper limit on both the vacuum and
single-photon fidelities for an XPM-based gate that produces uniform
$\PI$-radian phase shifts.

\section{Principal Mode Projection}

The fidelity upper limit we have found for both the vacuum and
single-photon fidelities increases with decreasing phase shift, so a
natural question arises: Can we cascade a series of small-phase-shift
gates, interspersed with quantum error correction, to realize a
high-fidelity {\sc cphase} gate? The errors addressed by
quantum-computation error correction---dephasing noise, depolarizing
noise, bit flips, etc.---all lie \emph{within} the Hilbert space for
the qubits of interest \cite{nielsen2000quantum}. In our case,
however, phase noise randomly distorts the single-photon pulse shape
while it preserves photon flux, $\Iin_K(t)=\Iout_K(t)$, so there is no
photon loss. Thus it causes the state to drift \emph{outside} the
computational Hilbert space, rendering traditional
quantum-error-correction techniques of no value.

An alternative approach would be to reshape the pulses after each XPM
interaction, but the random nature of the phase noise precludes this
approach's succeeding. Instead, let us pursue the route of
principal-mode projection (PMP), as suggested
in~\cite{PhysRevA.87.042325}. There, a $\vee$-type atomic system in a
one-sided cavity was part of a unit cell comprising the atomic
nonlinearity followed by filtering to project its output onto the
computational-basis temporal mode (the \em principal\/\rm\ mode).
Cascading a large number of these unit cells---each producing a small
phase shift but with an even smaller error---yielded the $\PI$-radian
phase shift needed for a {\sc cphase} gate with a fidelity that, in
principle, could be arbitrarily high if enough unit cells were
employed. It behooves us to see whether a similar favorable error
versus phase-shift tradeoff applies to our XPM system. Sadly, as we
now show, such is not the case.

Consider a single iteration of XPM+PMP when $\Ein_A(t)$ is in 
state $\alpha\ket{0}_A+\beta\ket{1}_A$,  with $|\alpha|^2 + |\beta|^2 = 1$, {and $\Ein_B(t)$ is in its vacuum state.  The density operator for $\Eout_A(t)$ will then be 
\begin{align}
  \hat{\rho}^{(0)}_{\text{PMP}} &= (1 - |\beta|^2\braket{|\hat{\mathcal{T}}|^2})
  \ket{0}_A\!\bra{0} + \alpha\beta^*\braket{\hat{\mathcal{T}}^\dagger}
  \ket{0}_A\!\bra{1} \nonumber \\ 
  &\,\,+ \alpha^*\beta\braket{\hat{\mathcal{T}}}\ket{1}_A\!\bra{0}
  + |\beta|^2 \braket{|\hat{\mathcal{T}}|^2}
  \ket{1}_A\!\bra{1},\label{eq:pmp-out0}
\end{align}
where $\hat{\mathcal{T}}\equiv\int \dd{t} |\psi_A(t)|^2
\ee^{\ii \hxi_A(t)}$ can be thought of as the photon-flux
transmissivity of the abstract pulse-shape filter responsible for
carrying out the PMP.   If the XPM interaction produces a uniform $\phi$-radian phase shift, then the \em same\/\rm\ expression gives the density operator for $\Eout_A(t)$ when $\Ein_B(t)$ is in its single-photon state $|1\rangle_B$.  Consequently, after averaging $\{\alpha, \beta\}$ over the Bloch sphere we find that the vacuum and single-photon fidelities satisfy
\begin{align}
  F_0  &= F_1 = \frac{1}{2} + \frac{1}{3}\braket{{\rm Re}(\hat{\mathcal{T}})} +
  \frac{1}{6}\braket{|\hat{\mathcal{T}}|^2}\label{eq:pmp-vac-fid}\\
  &= \frac{1}{2} + \frac{1}{3} \braket{\ee^{\ii \hxi_A(t)}} \nonumber \\[.05in]
  &\,\,+
  \frac{1}{6}\int \dd{t} \int \dd{s} |\psi_A(t)|^2 |\psi_A(s)|^2
  \braket{\ee^{\ii[\hxi_A(t)-\hxi_A(s)]}},
  \label{pmp-fid}
\end{align}
where we have used the fact that $\braket{\ee^{\ii \hxi_A(t)}}$ is
constant and real-valued. Comparing this result to
Eq.~(\ref{eq:vac-fid}), we see that a single iteration of PMP
\emph{does} increase both $F_0$ and $F_1$, but it does \em not\/\rm\
increase $F_\text{max}$ from what is given in (\ref{eq:vac-fid-max}),
a bound that still applies to both the vacuum and single-photon
fidelities.

Now it is easy to see that cascading $N$ unit cells of XPM+PMP cannot avoid the fidelity limit identified in the previous section.  For such a cascade, $F_0$ and $F_1$ obey Eq.~(\ref{eq:pmp-vac-fid}) with 
$\hat{\mathcal{T}}$ replaced by $\prod_{n=1}^N\hat{\mathcal{T}}_n$, where $\hat{\mathcal{T}}_n\equiv\int \dd{t} |\psi_A(t)|^2
\ee^{\ii \hxi_{A_n}(t)}$ is the
photon-flux transmissivity of the $n$th XPM+PMP unit cell. But, the $\{\hxi_{A_n}(t)\}$ are statistically independent and identically distributed, so that Eq.~(\ref{pmp-fid}) for the $N$ unit-cell cascade 
is then
\begin{align}
  F_0  &= F_1 = \frac{1}{2} + \frac{1}{3} \prod_{n=1}^N\braket{\ee^{\ii \hxi_{A_n}(t)}} \nonumber \\[.05in]
  &\,\,+
  \frac{1}{6}\int \dd{t} \int \dd{s} |\psi_A(t)|^2 |\psi_A(s)|^2
  \prod_{n=1}^N\braket{\ee^{\ii[\hxi_{A_n}(t)-\hxi_{A_n}(s)]}} \\[.05in]
  &\le \frac{2}{3} + \frac{1}{3}\exp\!\left(-(t_h/2) \int \dd{\omega}|H_\text{im}(\omega)|\right),
  \label{PMPbound}
\end{align}
where the inequality is obtained by assuming that each XPM+PMP unit
cell operates in the slow-response regime and provides a uniform phase
shift of $\PI/N$. That this fidelity bound coincides with
$F_\text{max}$ for a single XPM interaction that produces a uniform
$\PI$-radian phase shift is a consequence of quantum XPM's phase shift
and error scaling identically with the nonlinearity's strength,
$\eta$.

\section{Fiber-XPM Fidelity Bounds}
In this section we will evaluate the fidelity bound $F_\text{max}$ for two XPM response functions:  a reasonable theoretical model and the experimentally-measured response function of silica-core fiber.
We start with the family of single-resonance, two-pole
response functions characterized by the frequency response
\begin{align}\label{eq:response}
H(\omega) = \frac{\omega_0^2}{\omega_0^2 - \omega^2 - \ii \omega \gamma}.
\end{align}
This family, which was employed in~\cite{PhysRevA.73.062305}, includes a common approximation to the Raman response function of silica-core
fiber \cite{Lin:06}. For $0 < \gamma/2 < \omega_0$, its response
function $h(t)$ is underdamped,
\begin{align}
h(t)  = \frac{\omega_0^2\ee^{-\gamma t/2} \sin\left(\sqrt{\omega_0^2
    - \gamma^2/4}\,t \right)}{\sqrt{\omega_0^2 -
  \gamma^2/4}} \text{, for } t \geq 0;
\end{align}
for $\gamma/2=\omega_0$ it is critically damped,
\begin{align}
h(t) = \omega_0^2 t \ee^{-\omega_0t} \text{, for } t \geq 0;
\end{align}
and for $\gamma/2 > \omega_0$ it is overdamped,
\begin{align}
h(t) = \frac{\omega_0^2\ee^{-\gamma t/2} \sinh\left(\sqrt{\gamma^2/4 - \omega_0^2}\, t \right)}{\sqrt{\gamma^2/4 - \omega_0^2}} \text{, for } t \geq 0.
\end{align}
In all of these cases $h(t)$ has infinite duration, so we will optimistically take $t_h$
to be the root-mean-square duration of $h(t)$,
\begin{align}
  t_h & = \sqrt{\frac{\int_0^\infty \dd{t} t^2 h^2(t)}{\int_0^\infty
      \dd{t} h^2(t)} - \left(\frac{\int_0^\infty \dd{t} t
        h^2(t)}{\int_0^\infty \dd{t} h^2(t)} \right) ^2}\\
  & = \sqrt{\frac{1}{\gamma^2} + \frac{\gamma^2}{4\omega_0^4} -
    \frac{1}{2\omega_0^2}},
\end{align}
which satisfies
\begin{align}
\omega_0 t_h = \sqrt{\frac{1}{\Gamma^2} + \frac{\Gamma^2}{4} - \frac{1}{2}},
\end{align}
in terms of the dimensionless parameter $\Gamma = \gamma/\omega_0$.
This duration is minimized at $\Gamma = \sqrt{2}$, which is slightly
into the underdamped regime. In terms of $\Gamma$, it can be shown
that
\begin{align}\label{eq:arctan}
  \frac{\int \dd{\omega}|H_\text{im}(\omega)|}{\omega_0} & = \frac{\PI \ii + 2 \tanh
    ^{-1}\left(\frac{\Gamma ^2-2}{\Gamma \sqrt{\Gamma
          ^2-4}}\right)}{\sqrt{\Gamma ^2-4}},
\end{align}
which makes it easy to evaluate $F_\text{max}$, as a function of
$\Gamma$, from Eq.~(\ref{eq:vac-fid-max}), as shown in
Fig.~\ref{fig:fidelity}. Note that, despite its appearance, the
expression on the right in Eq.~(\ref{eq:arctan}) is real-valued for
$\Gamma\geq 0$.

\begin{figure}
\centering
\includegraphics[scale=0.3]{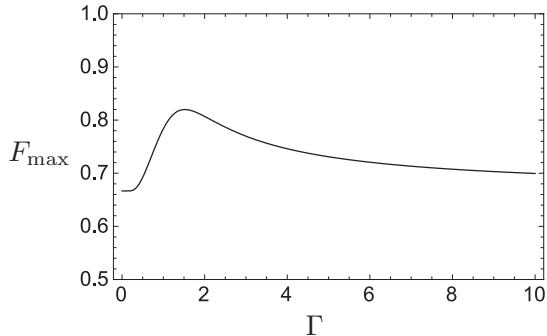}
\caption{Fidelity upper-bound $F_\text{max}$ for the single-resonance, two-pole response function plotted versus the normalized damping parameter $\Gamma$.}
\label{fig:fidelity}
\end{figure}

Figure~\ref{fig:fidelity} shows that $F_\text{max}$ peaks at just less than 82\%. It is worth emphasizing, in this regard, that $F_\text{max}$ is a very generous upper bound: (1) it does not include the effects of loss, dispersion, or self-phase modulation; (2) it assumes operation at $T=0$\,K; (3) it assumes operation in the slow-response regime, which would imply $\psi_A(t)$ and $\psi_B(t)$ had sub-femtosecond durations; (4) its use of $h(t)$'s root-mean-square duration for $t_h$ is an optimistic value insofar as uniform phase-shift conditions are concerned; and (5) it has generously set 
the third term of Eq.~(\ref{eq:vac-fid}) to its upper limit of 1.  Accordingly, it seems fair to say that, at least for this
response function, fiber XPM will not lead to a high-fidelity {\sc cphase} gate.

\begin{figure}
\centering
\subfloat[Frequency response.]{
\includegraphics[scale=0.35]{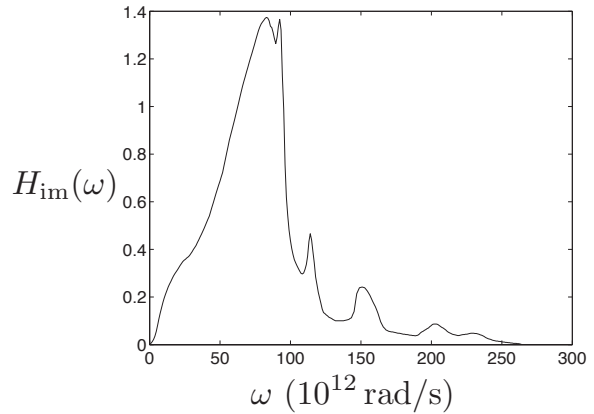}}\\
\subfloat[Temporal
response.]{
\includegraphics[scale=0.35]{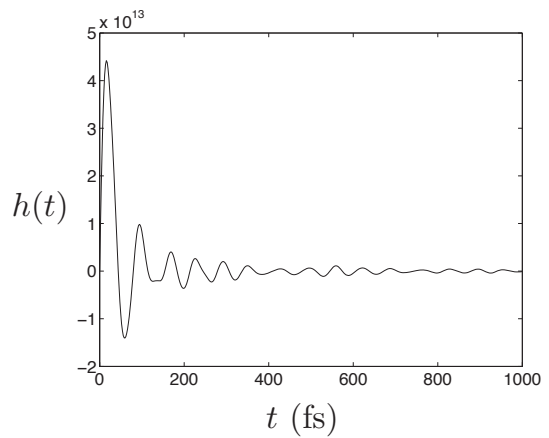}}
\caption{\label{fig:fiber-response} The Raman response of silica-core
  fiber, as measured by Stolen~et~al.~\cite{Stolen:89}.}
\end{figure}

At this point we could continue by evaluating the behavior of
$F_\text{max}$ for other idealized theoretical response functions, but
it is better to employ the XPM response function of fused-silica
fiber. XPM-based fiber-optical switches typically employ co-polarized
inputs \cite{974167}, and for such inputs that response function is
the fiber's co-polarized Raman response function \cite{Lin:06}, which
was measured by Stolen~\em et~al\/\rm.~\cite{Stolen:89} and is shown
in Fig.~\ref{fig:fiber-response}. For this response we have that the
root-mean-square duration of $h(t)$ is $t_h\approx 49.2\,\text{fs}$
and
\begin{align}
\int \dd{\omega}|H_\text{im}(\omega)| \approx 1.79\times 10^{14}\,\text{rad}/\text{s}.
\end{align}
These values imply that $F_\text{max}\approx 67.1\%$, which is worse than what we found for the $\Gamma$-optimized two-pole response.

If we stick with the Raman response function, we
can explore $F_1$ fidelity behavior when we relax our uniform phase-shift condition.
In particular, our uniform-phase-shift conditions make good sense when the
pulse width is significant relative to the response function's duration. However, deep in the slow-response regime---which these very same conditions suggest is optimal---the $\Ein_A(t)$ and $\Ein_B(t)$ pulse
shapes are well approximated by Dirac-delta distributions relative to the response function.  Thus it would
seem that ensuring the entirety of the pulse be exposed to the
entirety of the response is not particularly critical, in this regime, as there is
very little pulse to begin with. 

Suppose we are aiming for a $\PI$-radian phase shift.  Then, presuming operation at $T=0$\,K \em without\/\rm\ imposing the uniform phase-shift conditions, the vacuum and single-photon fidelities are bounded by (\ref{nonuniformF0bound}) for $F_0$, and 
\begin{align}
F_1 &\le F_1^\text{max} \equiv\frac{2}{3} - \frac{1}{3}
\exp\!\left(-\frac{\eta L}{4\PI}\int \dd{\omega} H_\text{im}(\omega)\right)\nonumber \\[.05in]
&\times {\rm Re}\left( \int \dd{t} \int
\dd{s} \ee^{\ii \eta\int^L_0\dd{z} h(t-s+z/u)}|\psi_A(t)|^2  |\psi_B(s)|^2 \right).
\label{nonuniformF1bound}
\end{align}
Decreasing the fiber length, $L$, at constant nonlinearity strength, $\eta$, mitigates the phase-noise fidelity degradation in $F_0$.  So long as $L$ satisfies the uniform phase-shift condition given in Eq.~(\ref{eq:uniform-phase-2}), $F_1$ will equal $F_0$, but once $L$ violates that condition we encounter a tradeoff for $F_1^\text{max}$ in Eq.~(\ref{nonuniformF1bound}):  the phase-noise factor, $\exp[-(\eta L/4\PI)\int \dd{\omega} H_\text{im}(\omega)]$, decreases with further decreases in $L$, but the factor it multiplies will be greater than the $-1$ value it had when the phase shift was uniform.  In Figs.~\ref{fig:f1-delta} and \ref{fig:f1-fast} we explore that tradeoff.

\begin{figure}
\centering
\includegraphics[scale=0.4]{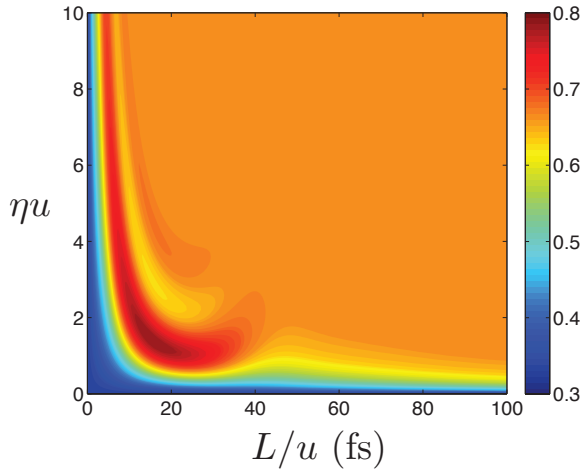}
\caption{\label{fig:f1-delta} (Color) Heat map of $F_1^\text{max}$ versus
  $\eta u$ and $L/u$ for Dirac-delta pulses with $t_d=L/2u$.}
\end{figure}

\begin{figure}
\centering
\includegraphics[scale=0.4]{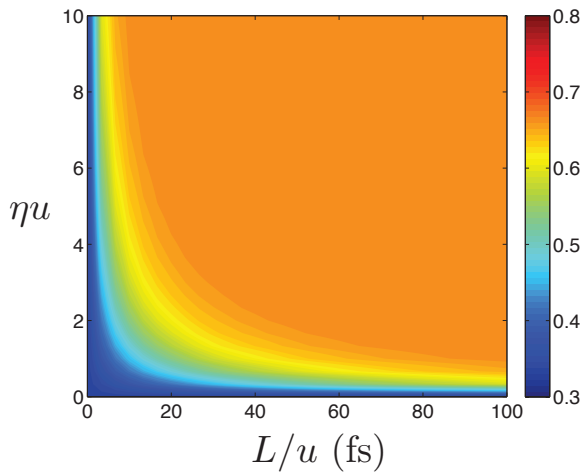}
\caption{\label{fig:f1-fast} (Color) Heat map of $F_1^\text{max}$ versus
  $\eta u$ and $L/u$ for Gaussian pulses with $t_\psi = 3\,\text{ps}$
  and $t_d=L/2u$.}
\end{figure}

Figure~\ref{fig:f1-delta} shows a heat map of $F_1^\text{max}$ 
as $\eta u$ and $L/u$ are varied.  Here we have assumed the extreme slow-response case of Dirac-delta pulses, and taken $t_d=L/2u$, so that the walk-off between the pulses is symmetric.  It turns out that $F_1^\text{max}$ peaks at approximately 78.6\% when $\eta u \approx 4.25$ and $L/u\approx
16\,\text{fs}$.  Although this peak value exceeds the 67.1\% $F_\text{max}$ value for fused-silica fiber, it is not very high and is lower than the optimum we gave earlier for the 
single-resonance, two-pole response function under uniform-phase-shift conditions. 
Figure~\ref{fig:f1-fast} shows a similar $F_1^\text{max}$ heat map for
3-ps-duration Gaussian pulses, i.e., $\psi_A(t) = \ee^{-2 t^2/t_\psi^2}/(\PI t_\psi^2/4)^{1/4}$ with $t_\psi = 3\,\text{ps}$. Here we see that the fidelity is abysmal, and our
numerical calculation does not yield an $F_1^\text{max} > 2/3$. As
expected, the uniform-phase-shift conditions are
important here---causing the fidelity to be tightly bounded by the
phase-noise alone---because operation is well into the fast-response regime.

Taken together, our fidelity bounds for the theoretical and measured response functions permit us to confidently say that 
XPM in silica-core fiber cannot promise a high-fidelity $\PI$-radian
{\sc cphase} gate, even under exceedingly idealistic assumptions.\\ 

\section{Conclusions}

We have presented a continuous-time, quantum theory for cross-phase
modulation with differing group velocities and have provided a
framework for evaluating the fidelity of using quantum XPM to
construct a {\sc cphase} gate. We found that perfect fidelity is
impossible, even in theory, owing to causality-induced phase noise
associated with Raman scattering in fused-silica fiber. For a
reasonable theoretical response function and the
experimentally-measured response function of silica-core fiber, we
found that XPM will not support a high-fidelity {\sc cphase} gate,
even under a collection of strictly favorable assumptions.  In particular, our analysis ignores loss, dispersion, and self-phase modulation.  Loss is especially pernicious, considering the length of fused-silica fiber needed for a single-photon pulse to create a $\PI$-radian phase shift on another such pulse.  

It is worth noting that the silica-core fiber response function we
studied is that for co-polarized pulses. The response function for
orthogonally-polarized pulses is much faster than---and 1/3 the strength of---its co-polarized counterpart, owing to its being mediated by an electronic interaction, as opposed to the Raman effect that is responsible for co-polarized XPM.    
We are not aware of any experimental characterization of the co-polarized response function. Nevertheless, the results in this paper suggest that it too will likely lead to low fidelity, so long as it is non-instantaneous, if for no other
reason than that its extreme speed will force operation in the less favorable fast-response regime.

Some final comments are now germane with respect to what potential
{\sc cphase} gates are \em not\/\rm, as yet, precluded by our
analysis. First, our results do not apply to XPM contained within a
larger interaction system, such as a cavity. Some recent results have
suggested that cavity-like systems may support a high-fidelity {\sc
  cphase} gate, despite noise
\cite{PhysRevA.87.042325,PhysRevLett.110.223901}. To date, however, no
one has studied the {\sc cphase}-gate fidelity afforded by cross-Kerr
effect XPM within a cavity. Finally, it is unclear to what extent if
at all our results apply to dark-state-polariton XPM in
electromagnetically-induced transparency (EIT). EIT theories usually
assume an instantaneous interaction, which is sometimes taken to be
nonlocal \cite{PhysRevA.83.053826}. In the physical world, however,
a phenomenon is rarely truly instantaneous, regardless of how good an
approximation that may be for various working theories. Our work
suggests that phase noise may be an issue for EIT if the response
function is not truly instantaneous. That aside, recent work has shown
that even instantaneous, nonlocal XPM is subject to the same
fidelity-degrading phase noise, with limited exceptions
\cite{Marzlin:10}. Together with the fact that EIT involves Raman
interactions \cite{PhysRevLett.84.5094}, which are ultimately
responsible for phase noise in co-polarized fiber XPM, this suggests
that these systems might have to contend with the sort of fidelity
issues presented here. Beyond that, other work has quantified additional
fidelity-limiting issues, which may be present in continuous-time XPM, that seem
likely to affect EIT systems \cite{banacloche, he1, he2}.

\section*{ACKNOWLEDGMENTS}
This research was supported by the DARPA Quantum Entanglement Science and Technology (QuEST) program and the NSF IGERT program Interdisciplinary Quantum Information Science and Engineering (iQuISE).  G.~P.~Agrawal graciously provided the Raman response data originally collected by R.~H.~Stolen.

%\bibliography{paper}

\begin{thebibliography}{26}
\bibitem[{foo({\natexlab{b}})}]{footnote4} \bibinfo{note}{Usually,
    $\eta_K(z,t)$ is taken to be $\eta I_J(z,t)$, for $J\neq K$, which
    reflects the assumption that the material response is
    instantaneous. Instead, we have employed non-instantaneous
    material responses.}
\bibitem[{foo({\natexlab{c}})}]{footnote1}
\bibinfo{note}{These field operators have the following nonzero commutators:
  $[\hB(z,\omega),\hB^\dagger(z',\omega')] =
  [\hC(z,\omega),\hC^\dagger(z',\omega')] =
  2\PI\delta(z-z')\delta(\omega-\omega')$.}
\bibitem[{foo({\natexlab{d}})}]{footnote2}
\bibinfo{note}{Here, and in all that follows, integrals without limits are
  taken over the interval $(-\infty,\infty)$.}
\end{thebibliography}

\newcommand{\noopsort}[1]{} \newcommand{\printfirst}[2]{#1}
  \newcommand{\singleletter}[1]{#1} \newcommand{\switchargs}[2]{#2#1}

\end{document}